# Equivalent Circuit Model Recognition of Electrochemical Impedance Spectroscopy via Machine Learning

Shan Zhu, Xinyang Sun, Yuxuan Wang, Naiqin Zhao, and Junwei Sha*

School of Materials Science and Engineering and Tianjin Key Laboratory of Composites and Functional Materials, Tianjin University, Tianjin 300350, China

*Email: zhushanatme@gmail.com, shajw@tju.edu.cn

## Abstract

Electrochemical impedance spectroscopy (EIS) is an effective method for studying the electrochemical systems. The interpretation of EIS is the biggest challenge in this technology, which requires reasonable modeling. However, the modeling of EIS is of great subjectivity, meaning that there may be several models to fit the same set of data. In order to overcome the uncertainty and triviality of human analysis, this research uses machine learning to carry out EIS pattern recognition. Raw EIS data and their equivalent circuit models were collected from the literature, and the support vector machine (SVM) was used to analyze these data. As the result, we addresses the classification of EIS and recognizing their equivalent circuit models with accuracies of up to 78%. This study demonstrates the great potential of machine learning in electrochemical researches.

model; Pattern recognition; Support vector machine

# Introduction

Electrochemical impedance spectroscopy (EIS) is to study the mechanism of electrode materials, solid electrolytes, conductive polymers and corrosion protection by measuring the change of impedance with sinusoidal frequency [1]. The fuzziness of interpretation is probably the biggest problem for EIS technology. Constructing equivalent circuit model is the most widely used method for EIS analysis [2]. In this method, the electrochemical system is regarded as an equivalent circuit, which is composed of basic components in series or parallel, such as resistance (R), capacitance (C) and constant phase element (Q). The structure of these equivalent circuit and the value of each element can be measured and fitted. Consequently, the details of the electrochemical systems and the properties of the electrode processes can be analyzed by using the electrochemical meaning of these components [1,2].

The model selection should reflect the practical significance, and the consequence discussion are based on the assumption that there is an accurate model. At present, the common method is screening out several potential equivalent circuit models according to the different applications, and then using mathematical fitting to simulate the corresponding pattern. Nevertheless, when several trusted models are available, they should be sorted to find the most reasonable model. Therefore, many subjective factors and judgments involved in the process of analyzing the equivalent circuit of EIS. How to choose the suitable equivalent circuit model is the critical step in the EIS technology.

Machine learning is a kind of algorithm which automatically analyzes and obtains the rule from the data and uses the rule to predict the unknown data [3,4]. The most appropriate choice of a variety of possible models is a typical classification problem. Classification is a common task in machine learning. To this regard, machine learning has a great application prospect in dealing with EIS analysis. Among various machine learning technologies, we choose the support vector machines (SVM) to deal with this classification problem and recognize the most suitable of EIS results.

The mechanism of SVM is that data points are regarded as p-dimensional vectors, and these points can be separated by (p-1)-dimensional hyperplanes. There may be many hyperplanes that can categorize the data. SVM can find the best hyperplane which maximizes the distance to the nearest data point on each side. SVM has been applied to pattern recognition (pattern recognition) problems such as portrait recognition (face recognition), text classification (text categorization) and so on.

In this paper, a SVM was constructed to analyze the suitable model of EIS results. The raw EIS data were collected from published articles, which focus on the electrochemical energy storage applications (i.e. batteries and supercapacitors). Hundreds of EIS data and their equivalent circuit models were studied by SVM.

# Methods

To conduct this research, the first step is to establish an EIS database, thus we extracted over 250 sets of EIS data and their equivalent circuit from published papers. There are two main reasons to extract data from the existing literature. First, there are a lot of related researches, suggesting that the total amount of data is large enough. Considering that the test instruments, test specifications and selected parameters used by each researcher are very different, it can provide enough universality for our data analysis process. On the other hand, in the published papers, the equivalent circuit models of EIS must have been carefully selected by the researchers, meaning that the author helping us to carry out an efficient model screening and improves the accuracy of the model as much as possible.

The way we extract data from the literature is to use the open source software WebPlotDigitizer for manual data extraction. Firstly, the original EIS data is extracted from the literature, and then the WebPlotDigitize tool is used to discretize the original EIS data into readable data points. The discretized data are in the form of two columns of data, i.e. Z' and -Z''. At the same time, to make the EIS data from various states comparable, we normalize the extracted data before conducting SVM.

The SVM were conducted on open source software (Python and Sklearn). Datasets are randomly divided into two categories: one is a training set; the other one is a test set. The data of the test set account for 20% of the total data.

# Results and Discussion

During the process, the normalized EIS is used as the input data and the corresponding equivalent circuit as the output data. The data analysis system is established by machine learning algorithm. The whole process is shown in Figure 1.

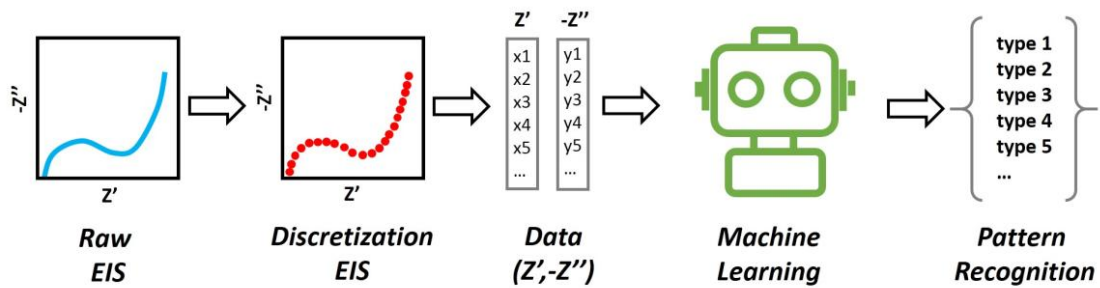

**Figure 1**. Process diagram of pattern recognition of EIS via machine learning.

After extracting enough EIS data, the next step is to classify the equivalent circuit of EIS. The EIS data we selected come from electrochemical energy storage devices, especially lithium-ion batteries and supercapacitors. 5 main equivalent circuit models are summarized as shown in Figure 2.

The reason why these five types is from the essence of EIS equivalent circuit. All these models consist of three components, which are impedance, constant phase element (CPE) and Warburg element (W). Impedance is a general term for the hindrance of resistance, inductance and capacitance. In the actual test, capacitors in EIS experiments often do not behave ideally. Instead they act like a constant phase element (CPE). Warburg element is used to describe the electrode behavior when the charge diffuses through a barrier layer. At very low frequencies, charged ions can spread deep and even penetrate the diffusion layer to produce a finite thickness Warburg element, which, if the diffusion layer is thick or dense enough, will result in even at a low limit of frequency, forming an infinite thickness of Warburg elements.

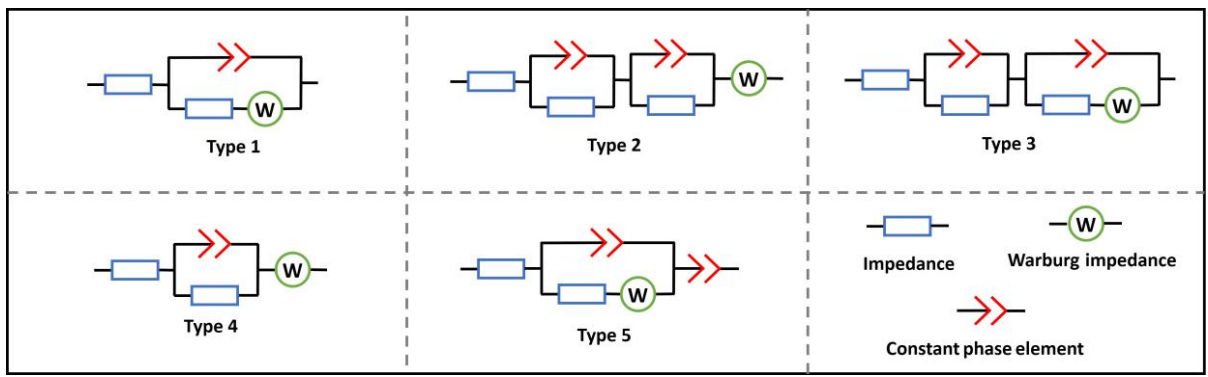

**Figure 2**. 5 different equivalent circuit models. The electrical elements represented by the various symbols listed below.

On the basis of obtaining 5 typical patterns, we list the representatives of several data under each type (Figure 3). From these data, it can be seen that under the same category, the spectrum of the data varies greatly; or the shapes of data are similar, yet they belong to different patterns. It is precisely because of this fuzziness of data classification that the traditional EIS analysis method has brought a lot of trouble.

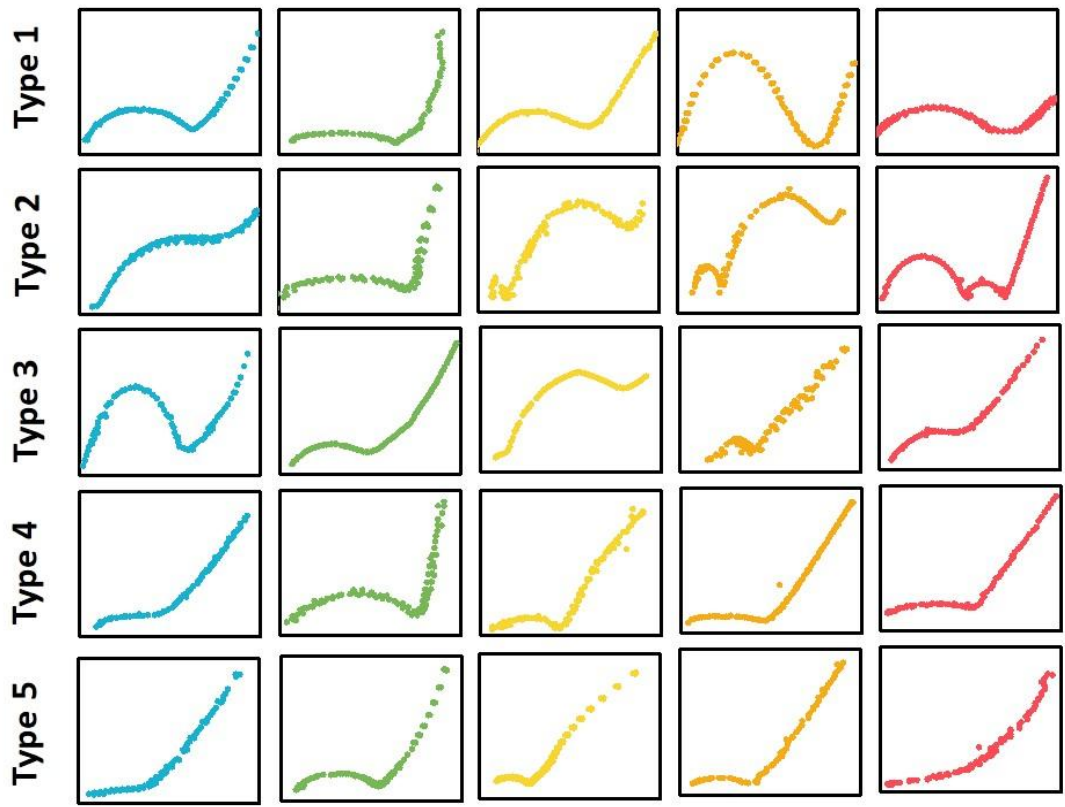

**Figure 3**. Examples of equivalent circuit for different patterns.

The first step of conducting SVM is to select the appropriate kernel functions. There are three kind of mainstream kernel functions which are "RBF", "Linear" and "Poly". By testing these kernel functions in the same data set, "Linear" achieves the best accuracy on the train set (86%). On the test set, the performance of "RBF" is best (78%). On balance, "RBF" was chossed as the kernel functions for the consequence process.

One of the most important parameters in the SVM is the penalty coefficient "C", that is, the tolerance to the error. The higher the C is, the more intolerable the error, meaning that SVM is easier to overfit. The smaller the C is, the easier it is to underfit. By testing of different C values, it is found that the overall perofrmance is best when C equal to 1. Gamma is an argument that comes with the RBF function after it is selected as the kernel. Implicitly determines the distribution of the data after mapping to the new feature space. The larger the gamma, the less the support vector; the smaller the gamma value, and the more the support vector. The number of support vectors affects the speed of training and prediction. With the increase of gamma, the accuracy of training set increases, but the accuracy of test set decreases, indicating that there is a tendency of over-fitting.

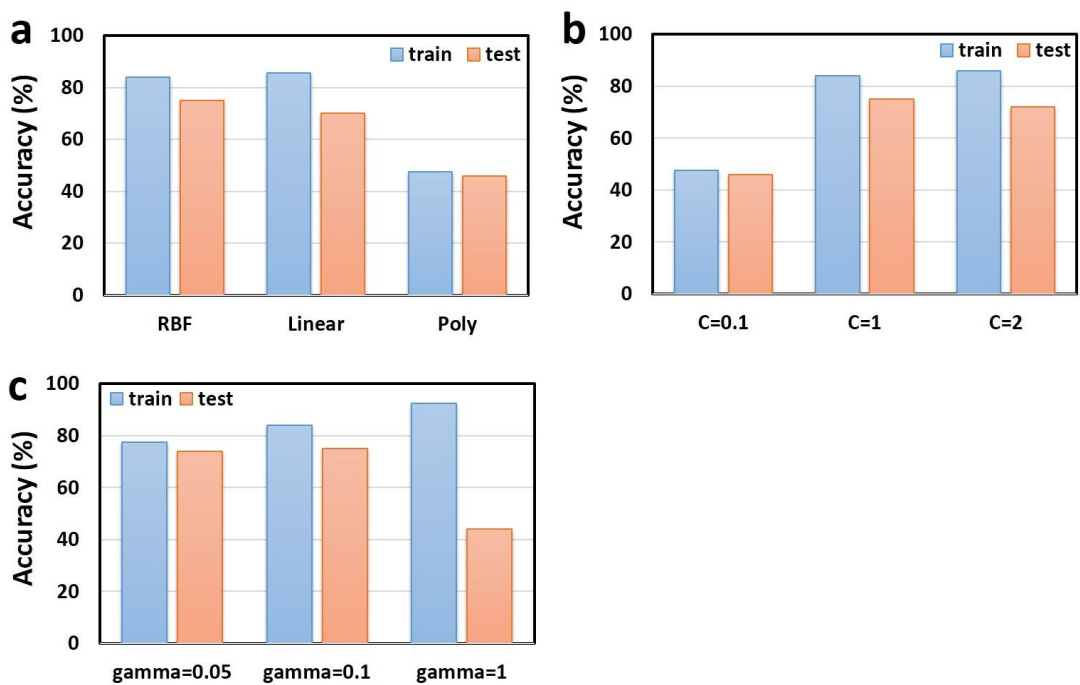

**Figure 4**.The influence of different parameters on the accuracy of SVM in predicting EIS model: (a) kernel functions, (b) penalty coefficient "C", (c) gamma value.

In sum, using RBF kernel function, when C is 1, gamma is 0.1, which can achieve good accuracy in train set (84%) and test set (78%). The errors of the experiment were mainly caused by the small number of data. Also, the EIS data used in this research come from electrochemical energy storage devices. In the future, we will expand the database to electrochemical corrosion, electrolytic water and other fields and to override as many EIS patterns as possible.

## Conclusion

This work applied SVM to deal with EIS and provide classification suggestions for EIS equivalent circuit model. A large number of EIS were extracted from published papers, and five typical equivalent circuit models were summarized. SVM using RBF kernel function achieves an acceptable prediction result (78%). More importantly, this work shows the great potential of machine learning electrochemical researches.

## Acknowledgement

This work was supported by the National Natural Science Foundation of China (Grant Nos. 51472177, 51772206 and 11474216). Thanks for the discussion from Prof. Jianrong Wang and Xiaoyang Gao.